\documentclass[aps,twocolumn,showpacs]{revtex4}
\usepackage{graphicx}
\usepackage{dcolumn}
\usepackage{bm}

\begin{document}
\preprint{J. Chem. Phys.}
\title{\bf 
Methanol-water solutions: a bi-percolating liquid mixture
}

\author{L. Dougan$^{(1)}$ } 
\author{S.P. Bates $^{(1)}$ }
\author{R. Hargreaves $^{(1)}$ }
\author{J. P. Fox $^{(1)}$ }
\author{J. Crain$^{(1)}$ $^{(2)}$ }
\author{J.L. Finney $^{(3)}$ } 
\author{V. R\'{e}at $^{(4)}$ } 
\author{A.K. Soper $^{(5)}$ } 

\affiliation{$^{(1)}$ School of Physics, The University of Edinburgh,
Mayfield Road, Edinburgh EH9 3JZ, UK}

\affiliation{$^{(2)}$ IBM TJ Watson Research Center, 1101 Kitchawan Road, Yorktown Heights, New York, 10598, USA}

\affiliation{$^{(3)}$ 
Department of Physics and Astronomy, University College London, Gower Street, London, WC1E 6BT, UK
}

\affiliation{$^{(4)}$ 
Institut de Pharmacologie et de Biologie Structurale, UMR 5089 - CNRS/UPS, Laboratorie ``RMN et interactions proteines-membranes'', 205 Route de Narbonne, 31077 Toulouse Cedex, FRANCE.
}
\affiliation{$^{(5)}$ 
ISIS Facility, Rutherford Appleton Laboratory, Chilton, Didcot, Oxon, OX11 OQX, UK
}

\date{\today}

\begin{abstract} 

An extensive series of neutron diffraction experiments and 
molecular dynamics simulations
has shown that mixtures of methanol and water exhibit
extended structures in solution despite the components being fully 
miscible in all proportions. Of particular interest is a
concentration region(methanol mole fraction between 0.27 and 0.54) where {\em both} methanol and water appear to form 
separate, percolating networks. This is the concentration range where many transport properties and 
thermodynamic excess functions reach extremal values. 
The observed concentration dependence of several of these material properties 
of the solution may therefore have a structural origin.  

\end{abstract}


\maketitle

\section{Introduction and motivation}

In aqueous solutions, amphiphiles 
show very rich and interesting behaviour
governed by the tendency of the molecules to self-organise into 
structures where the hydrophobic regions of molecules tend to be pushed together and away from the water, enabling the hydrophilic headgroups to hydrogen-bond more easily to the surrounding water molecules. This results in various 
supramolecular assemblies including micelles, 
columnar phases and lamellar structures 
depending on concentration and temperature.  An emerging 
route toward the development, testing and refinement of 
detailed molecular 
models of the hydrophobic interaction, hydration and the physics of 
aqueous macromolecules 
involves the use of small molecule systems (such as lower alcohols) 
as "prototypes". 

Despite their structural simplicity, 
it is well known that the thermodynamic and transport properties of
alcohol-water mixtures, such as the mean molar volume, the diffusion
coefficient, the compressibility and the excess entropy, are
significantly smaller, and the viscosity significantly larger, than
the values that might be expected from an ideal mixture of the pure
liquids\cite{gibson,lama,schott,yergovich,mcglashan,randzio}. 
The longstanding explanation of these effects in terms of
an enhanced structuring of the water in the presence of the 
alcohol\cite{frank} does
not appear to be supported  by modern diffraction 
experiments\cite{dixit1,dixit3,fs,finney105} and an alternative model is needed. Recent neutron diffraction studies of alcohol-water
binary mixtures\cite{dixit1,bowron1,bowron2,dixit3} 
are leading to new insights into the behaviour of
water near molecules containing both hydrophobic and hydrophilic groups. These have established that, in the dilute alcohol limit, 
the alcohol molecule appears to have a
mildly compressive effect on the water structure, as is seen from the
slight inwards movement of the second peak of the water-oxygen radial
distribution function compared to the same function in pure
water. This second peak, which occurs near $r \approx$ 4.5 $\AA$ 
in the $O_{W}O_{W}$
radial distribution function of pure water, has widely been interpreted
as the signature of the tetrahedral ordering in water. 
By contrast, in the opposite (concentrated alcohol) limit, 
the system segregates into what is effectively a
molecular-scale microemulsion\cite{dixit1}, 
with methyl head groups pushed towards
each other, and the hydrophilic hydroxyl groups forming a boundary around small pockets of a water-like fluid.  

These simple systems have also been the subject of considerable  
computational investigations. 
The earliest of these \cite{bolis8201,jorg8301, okaz8401} used 
Monte Carlo methodologies at low or infinitely dilute concentrations of
alcohol. Despite different 
computational models, and some apparent contradictions between 
their results, they all found an enhanced cagelike structure 
of water around the 
methyl group, in accordance with the Frank and Evans model\cite{frank}. 
Later, MD simulations explored other mixture 
compositions \cite{pali9101,ferr9001} using effective potential models. 
Tanaka and Gubbins \cite{tana9201} were amongst the first to highlight 
the role of the water-water interactions in discussing aqueous solutions.
More recently, Meng and Kollman \cite{koll9601} 
have performed 
MD simulations of various solutes (including methanol) at infinite 
dilution 
and 
found that the water structure around the hydrophobic groups is 
preserved rather than enhanced.
Laaksonen {\em et al} \cite{laak9701} have explored
several concentrations though there was no specific attention 
given to clustering. {\it Ab initio} simulations of alcohol-water
mixtures have also recently been reported \cite{meij0301,meij0302}, however
the computational expense of these simulations is such that they are restricted to 
picosecond simulation on small system sizes. Nonetheless, these studies have also 
pointed to the lack of structural enhancement of the water surrounding the hydrophobic moiety
in the alcohol. Very recently, some of us have obtained preliminary results from MD simulations of an alcohol-rich methanol-water solution that does exhibit extreme clustering and micro-immiscibility\cite{bate0401}

Given the ongoing interest in these systems\cite{sato,koga}, 
the availability of experimental data only at dilute 
and concentrated alcohol limits, and the apparently contradictory results from 
computer simulation
there is a strong motivation to undertake a 
systematic 
survey of extended structure (clustering) 
as a function of concentration in the model aqueous methanol system using both 
experimental and simulation techniques performed at identical state points. 
We are specifically interested in exploring the changes in the 
clustering behaviour as a function of concentration, and the extent to 
which molecular dynamics simulations account for the experimental 
observations and can provide additional insight into clustering dynamics.

\section{Experimental methods}
\subsection{Neutron diffraction experiments}
Protiated and deuteriated samples of methanol and water were obtained
from Sigma-Aldrich and used without additional purification. Neutron
diffraction measurements were performed on the SANDALS time-of-flight
diffractometer on the ISIS pulsed neutron source at the Rutherford
Appleton Laboratory, U.K. Samples were placed in flat plate cells made
from a Ti-Zr alloy that gives negligible coherent scattering. These were
mounted on a closed cycle refrigerator, and neutron diffraction
measurements were made at temperatures of 293K (with mole fraction $x=0.05$, $x=0.27$, $x=0.7$) and 298K
($x=0.54$) respectively. Corrections for attenuation and multiple
scattering were made using the ATLAS program suite. A further
correction for inelastic scattering was also made\cite{luz}. 
The differential
scattering cross-section for each sample was obtained by normalising
to a vanadium standard sample. A total of 7 samples were measured -
see Figure 1 for $x=0.54$. These were respectively (i) $CD_3OD$ in $D_2O$; 
(ii) $CD_3OH$ in
$H_2O$; (iii) a 50:50 mixture of (i) and (ii); (iv) $CH_3OD$ in $D_2O$; (v) a
50:50 mixture of (i) and (iv); (vi) $CH_3OH$ in $H_2O$; and (vii) a 50:50
mixture of (i) and (vi). For $x=0.05$ 5 samples were measured (i),(ii),(iii),(vi) and (vii).
These procedures  lead to  a structure factor  $F(Q)$ having  the form
$F(S_{\rm  HH}(Q),S_{\rm XH}(Q),S_{\rm  XX}(Q))$ 
where  S$_{HH}$(Q) gives
correlations  between labelled atoms  and S$_{XH}$(Q)  and S$_{XX}$(Q)
are  the  two  composite  partial  structure factors  which  give  the
remaining  correlations  between other  types  of  atoms  (X) and  the
labelled atom  type (H) in  the form of  a weighted sum  of individual
site-site partial structure factors.

\subsection{Empirical potential structure refinement}
A mixture of methanol and water contains 6 distinct
atomic components, namely C, O, M and H on the methanol molecule (here
M represents the methyl hydrogen atom while H represents the hydroxyl
hydrogen atom), and $O_{W}$ and $H_{W}$ on the water molecule. A full structural
characterisation of the system therefore requires the determination of
21 site-site radial distribution functions, which is well beyond the
possibility of any existing diffraction techniques by 
themselves. 

\begin{figure}
\centering
\includegraphics[angle=0,width=12cm]{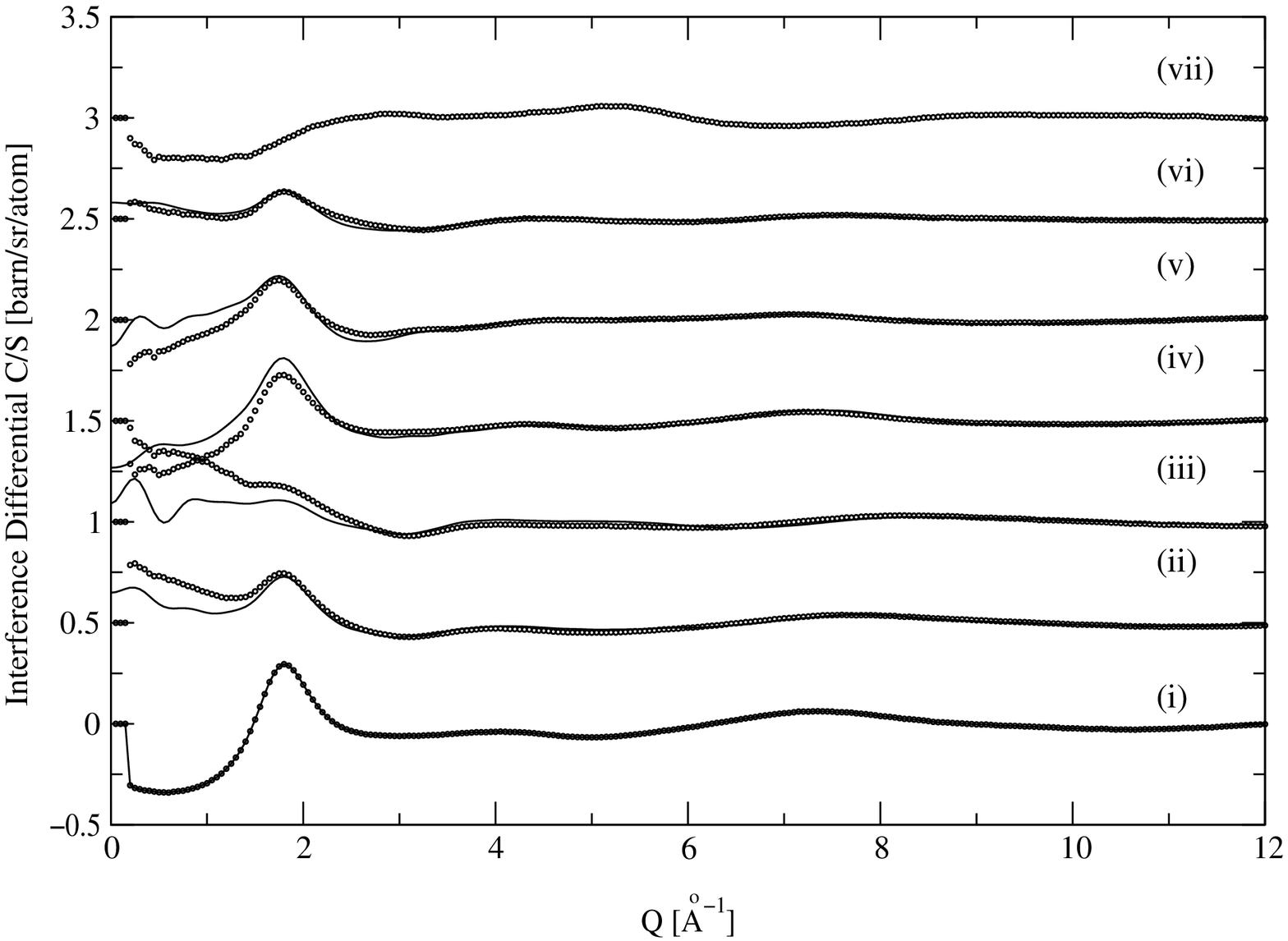}
\caption{Typical example of the fits (lines) obtained by the 
EPSR computer simulation procedure compared to the original data
(circles). The data shown in this case ($x = 0.54$) are the interference
differential scattering cross-sections for the samples (i) through
(vii) described under Methods. Discrepencies are observed in the low
$Q$ region. These are caused by difficulties in removing completely the effect
of nuclear recoil from the measured data. However this recoil effect
is expected to have only a monotonic dependence on $Q$ and so is
unlikely to influence the model structure to any significant extent.}
\label{fig:structure}
\end{figure}

Therefore to build a model of the liquid structure, 
the experimental data are used to constrain a computer
simulation of the mixture.  
However, unlike conventional simulations the 
empirical potential used here is obtained directly from the
diffraction data and has the effect of driving the structure of the three-dimensional model solution toward configurations that are consistent with the 
measured partial structure factors\cite{sop1}.
A total of 600 molecules(methanol and water) are
contained in a cubic box of the appropriate dimension to give the
measured density of each solution at the appropriate temperature(see Table I). Periodic boundary
conditions are imposed. Reference interatomic potentials for water and
methanol are taken from the literature\cite{methpot,watpot}. 
A comparison between the 
experimentally-measured partial structure factors and those generated from the 
ensemble-averaged EPSR configurations is shown in Fig \ref{fig:structure}.

In the present case a single set of site-site empirical potential coefficients was refined against the methanol-water data at each concentration, as well as for pure water\cite{soperA} and for pure methanol\cite{yama}. The result is a set of site-site empirical potentials which are consistent with methanol-water solutions over the full range of concentrations. Comparing the results of these simulations with those where the empirical potential coefficients are refined separately for each concentration revealed some discrepancies in the detail of the extracted $O_{W}-O_{W}$ radial distribution function, particularly at the higher methanol concentrations. Clearly the diffraction data by themselves do not constrain this function sufficiently to give a completely unambiguous $O_{W}-O_{W}$ radial distribution function. These discrepancies however do not affect the main conclusions of this paper, which are to do with the way methanol and water form distinct local clusters when mixed. We hope to present a more exhaustive study of the uncertainties involved in the EPSR analysis of molecular liquids and mixtures in a separate publication.

\begin{table}
\begin{center}
\begin{tabular}{|c|c|c|c|c|c|c|} \hline

Mole fraction & Temp. & Total No. & No. of methanol & No. of water & No. density & Box Size \\
$x$ & /$K$ & molecules & molecules & molecules & /  atoms/$\AA^{3}$ & /\AA \\
\hline
0.05 & 293 & 600 & 30 & 570  & 0.0995 & 26.68  \\ \hline
0.27 & 293 & 600 & 162 & 438 & 0.0967 & 28.69  \\ \hline
0.54 & 298 & 600 & 324 & 276 & 0.0955 & 30.73 \\ \hline
0.70 & 293 & 600 & 420 & 180 & 0.0930 & 32.04 \\ \hline

\end{tabular}
\end{center}
\caption{Parameters of the methanol-water mixtures used in the Empirical Potential Structural Refinement}

\end{table}

\section{Simulation and cluster analysis methods}

We have performed classical molecular dynamics simulations within the NVT ensemble, utilising 
previously tested intermolecular potentials for both methanol \cite{pere0101} and water
\cite{levi9701} that have been shown to predict the structure and dynamics of the 
single component liquids well. 
Both molecules are modelled as fully flexible entities, with explicit 
potential terms for each type of atom center. Using the {\tt DL\_POLY} code \cite{dlpoly}, we 
have performed simulations of 2ns duration with a timestep of 0.5fs on systems of $x$=0.27, $x$=0.54 and  $x$=0.7 mole fraction. The parameters relevant to these simulations are given in Table II.  All simulations were equilibrated for 0.5ns prior to 
data collection and trajectory snapshots were saved every 0.1ps for subsequent analysis. 

In analysing both the experimentally-constrained EPSR configurations and
molecular dynamics trajectories, an identical definition of a cluster is made 
based on 
bond connectivity.
Specifically, water molecules
are assigned to the same cluster if they can be connected by a
continuous hydrogen-bond network. The criterion used is that two 
water molecules are said to be hydrogen-bonded if their constituent oxygen atoms
are less than $R_{O_{W}O_{W}}$ apart, where $R_{O_{W}O_{W}}$ is determined as the minimum following the 
first peak in the O$_W$O$_W$ pair correlation function (approximately 3.5\AA\ for both
simulated and EPSR-fitted functions). 
For the methanol molecules, clusters may be defined in one of two ways. When investigating hydrogen-bonded clusters, we use the same criterion as for water, i.e. if constituent methanol oxygen atoms are less then $R_{OO}$ 
apart. However we also investigate clustering of methanol molecules via methyl group association. In this case, two methanol molecules are assigned
to the same cluster if the C-C distance is less than the minimum following the first peak determined from the CC pair correlation function (which is approximately
5.7 \AA). 
According to this criterion, methanol molecules that are
in contact only via their nonpolar groups are not within the 
same (hydrogen-bonded) cluster. 

When plotting cluster distribution histograms, we
plot the number of clusters of a size $i$, $m(i)$, as a fraction of the total 
number of clusters, $M$, where $M=\sum_i m(i)$ . 
Cluster lifetimes were deduced from the MD trajectories by calculating the average duration 
that a cluster persists with at least one participant member from the previous trajectory snapshot. 
The distinction is made between clusters of size 1 (i.e. with no H-bonded neighbours) and those of 
larger sizes. 

We have investigated the dependence of cluster distributions obtained on the size of the system employed. Using four different system sizes ranging from 700 to nearly 18000 particles, we find that the cluster distributions and lifetimes are practically identical in all cases and conclude that our observations are not significantly dependent on the choice of system size.


\begin{table}
\begin{center}
\begin{tabular}{|c|c|c|c|c|c|c|} \hline

Mole fraction & Temp.  & Total No. & No. methanol & No. water & No. density & Box Size \\
$x$ & /$K$ & molecules & molecules & molecules & / atoms/$\AA^{3}$ & /\AA \\
\hline
0.27 & 298 & 600 & 162 & 438 & 0.0968  & 28.69 \\ \hline
0.54 & 298 & 600 & 324 & 276 & 0.0953 & 30.75 \\ \hline
0.70 & 298 & 424 & 297 & 127 & 0.0934 & 28.50\\ \hline

\end{tabular}
\end{center}
\caption{Parameters of the methanol-water mixtures used in the Molecular Dynamics Simulations}

\end{table}

\section{Results and discussion}

\subsection{Local structure}

\begin{figure}
\centering
\includegraphics[angle=270,width=12cm]{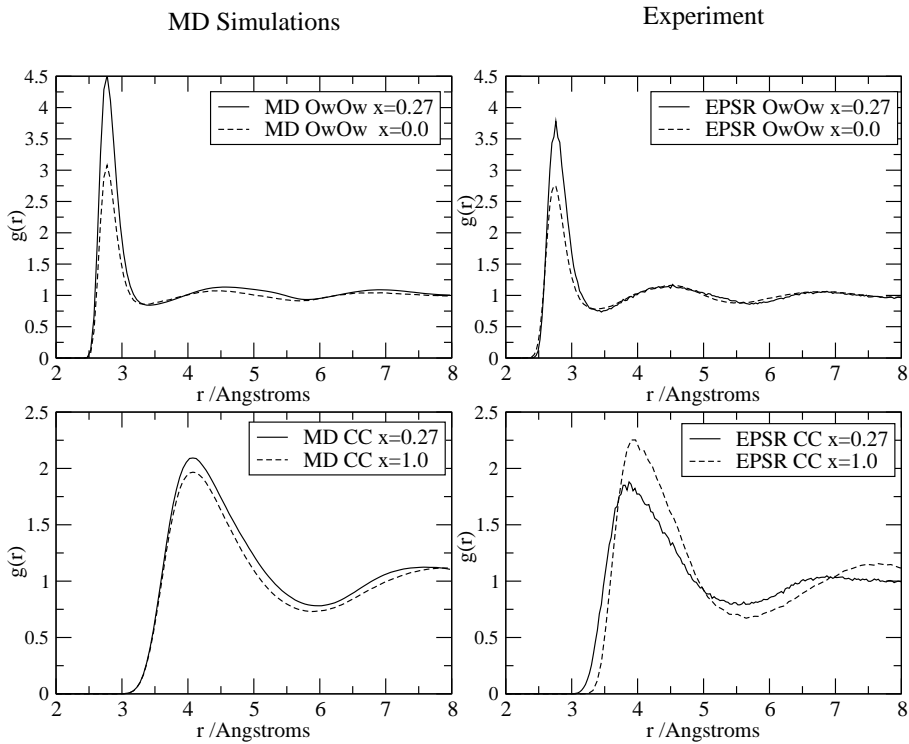}
\caption{
Comparison of the radial distribution function $g(r)$ as obtained from molecular dynamics($MD$) time averages(left) and
experimentally-constrained EPSR ensemble averages(right). The data shown is for the $x=0.27$ mixture in relation to the pure solvents, water($x=0.0$) and methanol($x=1.0$), for both the $O_{W}O_{W}$(top) and the C-C radial distribution functions(bottom).}
\label{simexpt_comparegr}
\end{figure}

A comparison of radial distribution fuctions $g(r)$ for $CC$ and $O_{W}O_{W}$ is shown in Fig \ref{simexpt_comparegr}.  A concentration of $x=0.27$ and the pure solvents(water $x=0.0$ and methanol $x=1.0$) are shown for both the experimentally-constrained EPSR configurations and the 
molecular dynamics simulations.
Local structure therefore appears similar in both the experimental 
and computational results. There are subtle differences between the experiment and simulation particularly the $g(r)$ for $CC$ which could be due to different potentials used in the fitting procedure and the different treatments of methyl group flexibility. However, it is clear that there is good qualitative agreement when we go on to explore extended structures in the solutions. 

\subsection{Clustering and cluster lifetime}
\begin{figure}
\centering
\includegraphics[angle=0,width=9cm]{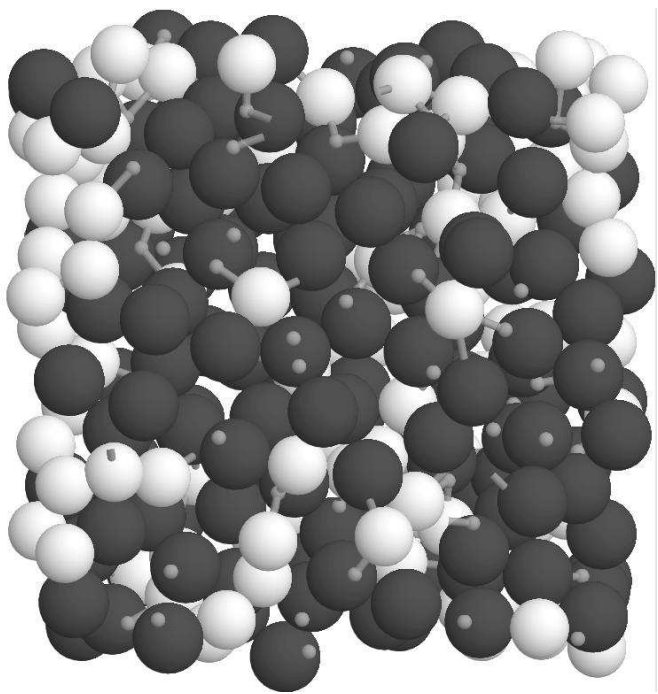}
\caption{Snapshot of an experimentally-constrained EPSR model of the methanol-water mixture at $x=0.54$ showing clusters of the segregated components. Methyl groups are shown as black spheres, large white spheres highlight the position of water molecules and small grey spheres denote methanol oxygen atoms.}
\label{expt_cluster}
\end{figure}

We consider first the results of the neutron diffraction experiments. 
Visual inspection of the boxes of atoms reveals significant
segregation of water from methanol at all concentrations. An example is 
shown for $x=0.54$ in Fig \ref{expt_cluster}. Visual inspection also 
suggests that the methanol clusters do not 
tend to form hydrogen-bonded chains to the same extent as in
the pure alcohol. Instead the methyl headgroups tend to be in contact,
with the hydroxyl headgroups bonding to water molecules forming the
main boundary between methanol- and water-rich regions. This is broadly as expected for a hydrophobically-driven system and is what has been observed in earlier diffraction work\cite{bowron1,bowron2}. Similar snapshots are 
obtained from 
the molecular dynamics simulations(not shown). 

Further evidence of the way in which the presence of water affects the structure and orientation of methanol molecules is obtained from investigation of the size of hydrogen-bonded methanol clusters from the MD simulations and radial distribution functions. Compared to the hydrogen-bonded
network in the pure liquid, the cluster sizes are much smaller. For example, in the methanol-rich solution ($x$=0.7) we find about $\sim$ 75 $\%$ of the hydrogen-bonded methanol clusters consist of only 1,2 or 3 methanol molecules. For comparison, in pure methanol the fraction of clusters of the same size range drops to around $\sim$ 37 $\%$, with the majority of hydrogen-bonded clusters containing more than 3 methanol molecules\cite{bate0401}. This indicates a substantial disruption to the methanol hydrogen-bonded network. The orientation of the methanol molecules is also affected; in the solution the methyl headgroups are pushed closer together, as evidenced by a shift to smaller $r$ values of the first peaks in $g_{CC}$ and $g_{MM}$, where $M$ denotes a methyl hydrogen, in both the EPSR procedure and MD simulations.

The persistence of clustered structures in these systems is reflected in 
the average lifetime of clusters and single molecules. An estimate of these lifetimes can be obtained from analysis of the molecular dynamics 
trajectories and are particularly interesting in the 
simulation 
performed at the  $x=0.7$ methanol mole fraction. Firstly, we observe 
that individual water molecules are short-lived and survive, on average, 
for only 2 ps before being absorbed into a cluster. However, 
in rare cases, lifetimes of 100 ps are found. A similar 
result is found for the other cluster sizes which show average lifetimes 
of about 3ps though there are also persistent clusters surviving 
for up to 0.5 ns. The methanol hydrogen-bonded clusters, already noted to be 
much reduced in size by the presence of the water, are extremely short-lived; most persist 
for approximately 1ps with no methanol hydrogen-bonded structure lasting for more than 40ps. 
Thus the simulation results, in addition 
to being consistent with the overall structures implied 
by the neutron diffraction data, also 
suggest that the extended structures characterising 
the methanol-water system are very dynamic with rapid shedding and 
reforming of cluster members. 

\subsection{Percolation}
We now explore clustering of both species quantitatively as a 
function of concentration where an identical definition of a cluster is made in analysing both the experimentally-constrained EPSR configurations and 
molecular dynamics trajectories. For water molecules the hydrogen-bond definition was used to designate which molecules belong to a given water cluster, while for methanol clusters the C-C distance definition was used, as this criterion is more indicative of the nature of the methanol clustering than the hydrogen bond criterion.

\begin{figure}
\centering
\includegraphics[angle=0,width=12cm]{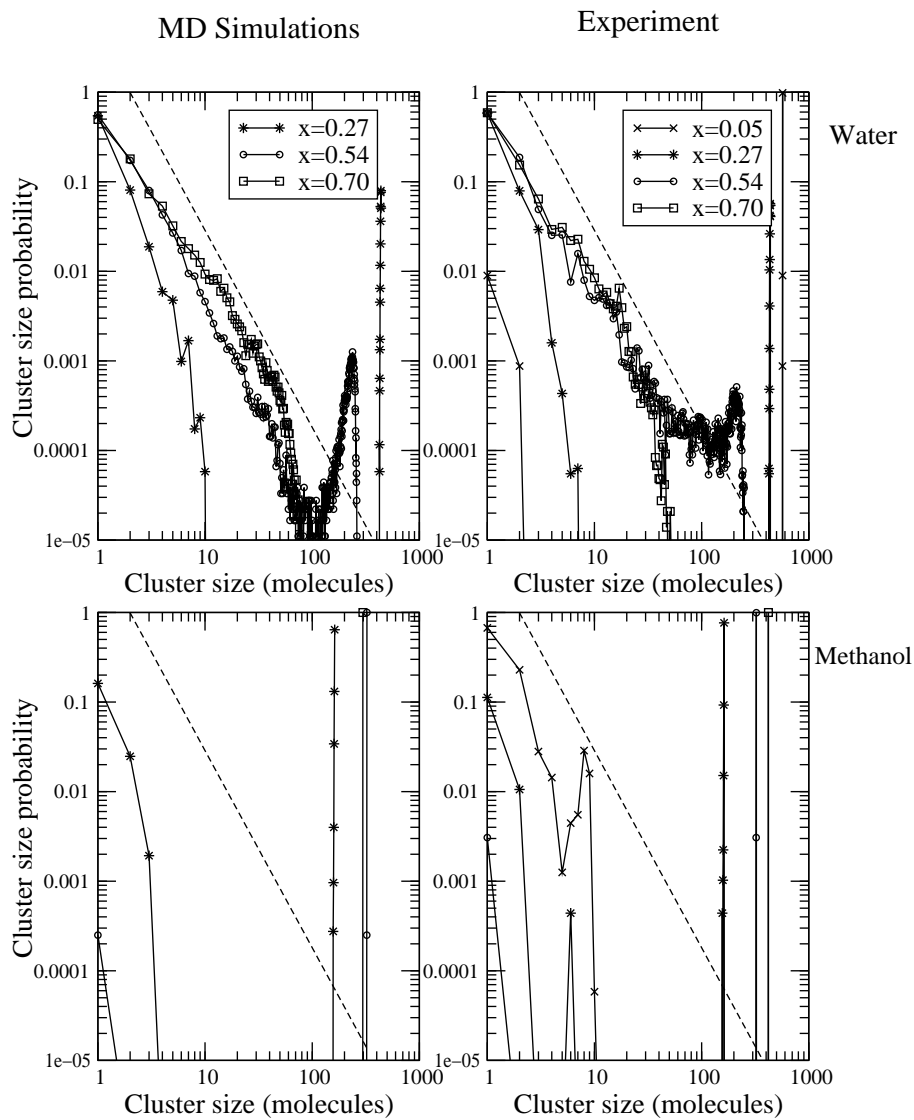}
\caption{Cluster size distributions for water(top) and methanol(below) clusters in methanol-water mixtures. For water molecules the hydrogen-bond definition was used to designate which molecules belong to a given water cluster, while for methanol clusters the C-C distance definition was used. On the left, from MD simulations with methanol mole fractions $0.27$ , 
$0.54$  and $0.7$ and on the right from neutron diffraction experiments for methanol mole fractions $0.05$, $0.27$ , $0.54$  and $0.7$  The dashed lines show the predicted cluster size distribution at the percolation threshold\cite{jan}. Percolation in the simulated box occurs when clusters of a size close to the number of molecules in the simulation box form (vertical lines on the right hand side of the plot).}
\label{fig:exp_sim_clusters}
\end{figure}

The cluster size distributions as obtained from the EPSR ensembles
(for $x= 0.7, 0.54, 0.27 \mbox{ and } 0.05 $  ) 
and molecular dynamics simulation (for $x= 0.7, 0.54 \mbox{ and } 0.27 $  ) are shown in Fig \ref{fig:exp_sim_clusters} along with the  predicted power law $n_s \approx s^{- 2.2}$ for  
random percolation on a 3-d cubic lattice\cite{jan}. The experimental and 
computational results both give similar results for the cluster size 
distribution at the three concentrations for which a direct comparison can be made 
as was implied by visual inspection of the EPSR and molecular dynamics 
structures.  

Several ``special'' concentrations emerge as defined by changes in clustering 
behaviour.
Specifically,  $x \approx 0.27$ determines the approximate 
alcohol concentration below which water percolates 
throughout the
mixture while methanol does not - 
occurring instead only in small, isolated clusters. Above
this molar fraction, however,  methanol percolates throughout the
mixture. The larger water clusters also percolate
(at $x=0.54$, see Fig.\ref{fig:exp_sim_clusters} ) 
but become increasingly isolated until they are confined to distinct, 
non-spanning clusters by $x = 0.7$. 
Thus, according both to the experimentally-constrained EPSR data and the MD simulations, in 
the approximate concentration range defined by $0.27 < x < 0.54$ 
{\em both}
water and alcohol clusters percolate simultaneously, making this a
bi-percolating liquid mixture.

Significantly, the mole fraction range 
over which simultaneous two-component percolation occurs 
coincides
closely with the concentration at which many thermodynamic properties
show extrema\cite{GibsonRE,LamaRF,Tomaszkiewicz,SchottH,Derlacki,Soliman}. This suggests that the nature and extent of 
clustering in these mixtures may offer a structural explanation for the 
thermodynamic anomalies. 

In earlier work on liquids, computer simulations 
have identified percolation transitions in supercritical 
Lennard-Jones fluids\cite{yoshii},
supercritical water\cite{kalinichev}, water in aqueous 
acetonitrile\cite{bergman} and aqueous tetrahydrofuran (THF) 
\cite{oleinikova,brovchenko2}.
However, to our knowledge this is the first report of simultaneous 
percolation of two fully miscible fluids. Despite subtle 
variation in the cluster size distribution, 
we have obtained independent evidence of two-component percolation 
from experimental and computational 
methods.  

\subsection{Dimensionality of clusters}
\begin{figure}
\centering
\includegraphics[angle=0,width=7cm]{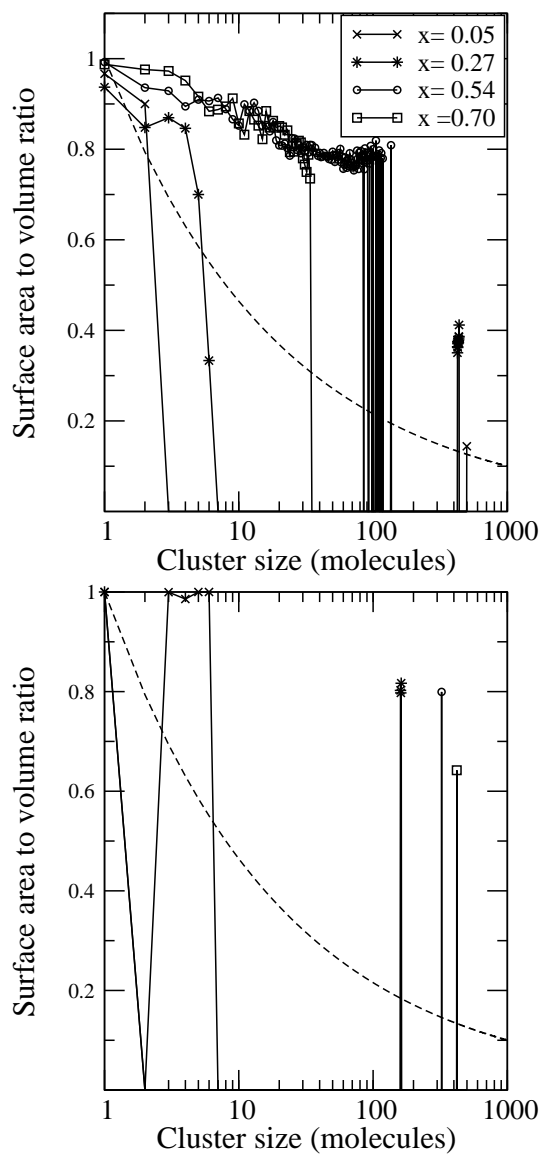}
\caption {Ratio of number of water molecules at the surface of a cluster (as defined by being hydrogen bonded to a methanol hydroxyl group) to total number of water molecules in a cluster(top). The dashed line shows the $N^{-1/3}$ behaviour that would be expected for this ratio if the clusters grew equally in 3 dimensions with N the number of molecules in a cluster. Only for the fully percolating water cluster at x = 0.05 do the clusters show normal 3D behaviour. Ratio of number of methanol molecules at the surface of a cluster to total number of methanol molecules in a cluster(below). Here even at x = 0.7 the methanol clusters do not approach full 3D behaviour.}
\label{fig:exp_dimension}
\end{figure}

A further feature of the EPSR clusters is revealed in 
Fig \ref{fig:exp_dimension}
which shows the typical surfacWe now explore clustering of both species quantitatively as a 
function of concentration where an identical definition of a cluster is made in analysing both the experimentally-constrained EPSR configurations and 
molecular dynamics trajectories. For water molecules the hydrogen-bond definition was used to designate which molecules belong to a given water cluster, while for methanol clusters the C-C distance definition was used, as this criterion is more indicative of the nature of the methanol clustering than the hydrogen bond criterion.

The cluster size distributions as obtained from the EPSR ensembles
(for $x= 0.7, 0.54, 0.27 \mbox{ and } 0.05 $  ) 
and molecular dynamics simulation (for $x= 0.7, 0.54 \mbox{ and } 0.27 $  ) are shown in Fig \ref{fig:exp_sim_clusters} along with the  predicted power law $n_s \approx s^{- 2.2}$ for  
random percolation on a 3-d cubic lattice\cite{jan}. The experimental and 
computational results both give similar results for the cluster size 
distribution at the three concentrations for which a direct comparison can be made 
as was implied by visual inspection of the EPSR and molecular dynamics 
structures.  

Several ``special'' concentrations emerge as defined by changes in clustering 
behaviour.
Specifically,  $x \approx 0.27$ determines the approximate 
alcohol concentration below which water percolates 
throughout the
mixture while methanol does not - 
occurring instead only in small, isolated clusters. Above
this molar fraction, however,  methanol percolates throughout the
mixture. The larger water clusters also percolate
(at $x=0.54$, see Fig.\ref{fig:exp_sim_clusters} ) 
but become increasingly isolated until they are confined to distinct, 
non-spanning clusters by $x = 0.7$. 
Thus, according both to the experimentally-constrained EPSR data and the MD simulations, in 
the approximate concentration range defined by $0.27 < x < 0.54$ 
{\em both}
water and alcohol clusters percolate simultaneously, making this a
bi-percolating liquid mixture.

Significantly, the mole fraction range 
over which simultaneous two-component percolation occurs 
coincides
closely with the concentration at which many thermodynamic properties
show extrema\cite{GibsonRE,LamaRF,Tomaszkiewicz,SchottH,Derlacki,Soliman}. This suggests that the nature and extent of 
clustering in these mixtures may offer a structural explanation for the 
thermodynamic anomalies. 

In earlier work on liquids, computer simulations 
have identified percolation transitions in supercritical 
Lennard-Jones fluids\cite{yoshii},
supercritical water\cite{kalinichev}, water in aqueous 
acetonitrile\cite{bergman} and aqueous tetrahydrofuran (THF) 
\cite{oleinikova,brovchenko2}.
However, to our knowledge this is the first report of simultaneous 
percolation of two fully miscible fluids. Despite subtle 
variation in the cluster size distribution, 
we have obtained independent evidence of two-component percolation 
from experimental and computational 
methods.  
e area to volume ratio (as represented
by the number of water molecules in a cluster which form hydrogen
bonds with a methanol molecule divided by the number of water
molecules in the cluster) for water molecules prior to the percolation
transition, together with an exactly analogous quantity for the methanol
clusters. Clearly the ratio does not decay as $N^{-1/3}$ as would be
expected for a 3-dimensional object: the clusters appear to maximise their surface area by forming as many bonds as possible
with methanol. The observed behaviour corresponds much more closely to
a 2-dimensional object, suggesting the clusters occur in the form of
disordered sheets or cylinders, rather than the sphere-like objects
that might be expected in conventional micelle formation. Only at
the highest water content, $x = 0.05$, do the water clusters appear to
have adopted 3D characteristics.  

\begin{figure}
\centering
\includegraphics[angle=270,width=9cm]{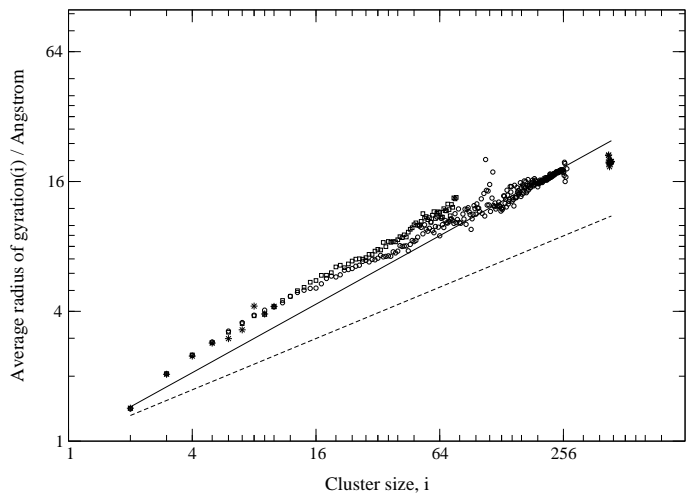}
\caption{
Average radius of gyration of all water clusters of size $i$ as a function of cluster size $i$, in solutions of different composition from MD simulations. The composition of the solutions, in terms of mole fractions of methanol, are x=0.7 (squares), 0.54 (circles), 0.3 (stars). The solid and dashed lines represent the variation of average radius of gyration with cluster size of percolating clusters on 2-d square lattice (characterised by $d$=91/48) and 3-d cubic lattice (characterised by $d$=2.53), respectively, as determined from large lattice random site percolation simulations\cite{jan}.}
\label{fig:fractals}
\end{figure}

This broad conclusion is also supported by the analysis of the topologies of the water clusters predicted from the MD simulations. Figure \ref{fig:fractals} shows the variation of the average radius of gyration of all the clusters of size $i$ ($R(i)$) as a function of cluster size, for three different solution compositions. The cluster topology can be characterised by a fractal dimension $d$, determined by a power law fit to the data in Figure \ref{fig:fractals} , such that $R(i) \propto i^{-d}$. For solutions of methanol mole fractions of $x$=0.7, 0.54 and 0.3, these values of $d$ are determined to be 1.69, 1.89 and 2.03, respectively. We note that as the mole fraction of water (and hence the proportion and size of larger clusters) increases the characteristic dimension increases. Interestingly, the cluster topologies of small clusters determined from our simulations are insensitive to changes in composition; all three compositions exhibit similar fractal dimension ($d\sim$ 1.6) for clusters sizes up to $i \sim$ 20 .
These results indicating the 2-d fractal dimension of water clusters over a range of compositions is in agreement with the results 
of a simulation study by Oleinikova {\it et al} \cite{oleinikova} on the percolation of water clusters in the vicinity of a region of 
immiscibility in an aqueous solution of THF. Simulataneous bi-percolation of both THF and water was found in this study, with
THF percolating clusters having a characteristic fractal dimension of 2.5 and those of water approximately 1.9. Visual inspection of some 
of our percolating clusters (of both methanol and water) reveal that these clusters span all six faces of the simulation box.

\section{Summary and conclusion}

Isotope-labelled neutron diffraction measurements  
analysed using the empirical potential structural refinment method
have been combined and compared with independent molecular dynamics 
simulations at identical state points to explore 
structure in methanol-water mixtures at several concentrations. We find 
that local and extended structures are well described 
by both methods and lead to 
similar conclusions. We find highly heterogeneous mixing across the entire concentration range despite apparent miscibility of both components in all proportions. Extended chain, sheet and three-dimensional structures form depending on concentration. 

At a particular concentration 
regime near $x=0.27$ these structures form percolating 
networks for both components. This concentration has long been 
considered "special" as it is near the point where many transport coefficients and 
thermodynamic functions have extremal values. Other alcohols also show 
extrema of these same material properties (at lower mole fractions) 
and the present work suggests a structural basis for these observations connected to 
the details of mixing heterogeneities.

\section*{Acknowledgements}
We are grateful to I. Brovchenko and G. Martyna for very helpful discussions

\bibliography{mwjcp6}
\clearpage

\end{document}